# Current controlled non-hysteresis magnetic switching in the absence of magnetic field

Yanru Li[1,2], Meiyin Yang[1,2]*, Guoqiang Yu[3,4], Baoshan Cui[3,4], and Jun Luo[1,2]*

[1]*Key Laboratory of Microelectronic Devices and Integrated Technology, Institute of Microelectronics, Chinese Academy of Sciences (IMECAS), Beijing 100029, China*

[2]*University of Chinese Academy of Sciences (UCAS), Beijing 100049, China*

[3]*Songshan Lake Materials Laboratory, Dongguan, Guangdong 523808, China*

[4]*Institute of Physics, Chinese Academy of Sciences, Beijing 100190, China*

E-mail: luojun@ime.ac.cn, yangmeiyin@ime.ac.cn

By means of local ion implantation, we investigated the influence of lateral interface on current induced magnetic switching by spin-orbit torque in a perpendicularly magnetized Pt/Co/Ta multilayer. The experimental results show that, in this system, the domain wall motion under electrical current can be affected by two mechanisms: symmetry breaking and current-driven Néel wall motion at the lateral interface. The dominant mechanism is symmetry breaking (current-driven Néel wall motion) at the large (small) current. Due to the competitive relationship of these two mechanisms, the non-hysteresis effect magnetic switching without an external magnetic field is obtained. Based on the non-hysteresis effect magnetic switching, we can realize AND and OR logic gates without resetting.

Spin-orbit torque (SOT) generated from spin currents offers an effective means to manipulate the magnetization of the ferromagnetic layer in the heavy-metal/ferromagnetic (HM/FM) bilayers[1-10]. In recent years, the SOT device with lateral interfaces has attracted remarkable attention due to its significance in logic functions realized by the current-induced DW motion[11-13]. Lateral interfaces break the symmetry of the system, leading to a gradient-induced additional spin-torque[14-16] to realize field-free deterministic switching. The origin of the gradient-induced additional spin-torque is proposed from the Rashba effect[17], asymmetric spin current[18], tilted anisotropy[19] and Néel chiral symmetry breaking[20]. Besides, in the system with lateral interfaces, current-driven Néel wall motion is observed when Dzyaloshinskii-Moriya interaction (DMI) is large, which enables electrically controlled of DW position without an external magnetic field[11]. Therefore, in a large DMI system with the lateral interface, both the symmetry breaking and current-driven Néel wall motion mechanisms have a great influence on the SOT-induced magnetic switching.

The contributions of these two mechanisms can be analyzed by effective fields $H_{eff(S)}$ and $H_{DL}$, which are originated from the symmetry breaking and current-driven Néel wall motion, respectively. For example, the $H_{eff(S)}$ can be generated from the symmetry breaking by magnetic anisotropy field gradient ($\nabla H_K$) (*ref* 20). And the $H_{DL}$ is due to the damping-like effective field ($M \times \sigma$, $\sigma$ is the spin vector) at the Néel wall by SOT[16]. The directions of $H_{eff(S)}$ and $H_{DL}$ can be the same or opposite. Under the positive current, when the magnetization (*M*) in the domain wall (DW) is pointed to –*x* (+*x*) under the symmetry breaking $\nabla H_K > 0$ ($\nabla H_K < 0$), $H_{eff(S)}$ and $H_{DL}$ are in the same direction, as illustrated in Fig. 1(a) and Fig. 1(d). When the *M* in the wall is along with +*x* (−*x*) under $\nabla H_K > 0$ ($\nabla H_K < 0$), $H_{eff(S)}$ and $H_{DL}$ point opposite, as shown in Fig. 1(b) and Fig. 1(c). In this case, there are competitions between $H_{eff(S)}$ and $H_{DL}$, and the magnetic switching will depend on the larger one of these two fields. But $H_{eff(S)}$ and $H_{DL}$ do not always exist simultaneously. For example, when the lateral interface is parallel to the current direction, $H_{eff(S)}$ plays a major role in deterministic magnetic switching induced by the SOT effect[14], rather than $H_{DL}$. The reason is that the *M* in the wall at the interface is parallel to the spin direction, so that $H_{DL} \approx 0$ ($M \times \sigma \approx 0$). On the other hand, when the lateral interface is perpendicular to the current direction, maximum $H_{DL}$ can be produced[21, 22], because *M* in the wall is perpendicular to the direction of σ. In this case, $H_{eff(S)}$ almost vanishes and hardly contributes to the magnetic switching at this interface[19]. So, adjusting the angle between lateral interface and current direction can control the contribution of symmetry breaking and current-driven Néel wall motion mechanisms to magnetic switching. However, the research on the competition between symmetry breaking and current-driven Néel wall motion is still lacking when these two mechanisms contribute to magnetic switching simultaneously.

In this work, we study the contribution of symmetry breaking and current-driven Néel wall motion mechanisms on field-free magnetic switching by designing the angle between lateral interface and current direction. The lateral interface is created by the ion implantation at the center of the Hall device. The angle of the interface is 45° with the current direction ($x$ axis). Theoretically, this 45-degree lateral interface can be decomposed into a 0-degree and a 90-degree interface, so that both symmetry breaking and current-driven Néel wall motion have a great contribution to magnetic switching. We create a competitive scenario of these two mechanisms the same as one in Fig. 1(b) for this device. The results indicate that the symmetry breaking (current-driven Néel wall motion) mechanism plays a major role at a large (small) current pulse, which results in the non-hysteresis effect switching by electrical current amplitude. Using the non-hysteresis effect magnetic switching, we can realize AND and OR logic gates without resetting before each logic operation. Compared with logic gates with resetting, the non-resetting AND and OR logic gates can save time, area and power.

The samples with the structure of Ta (2 nm)/Pt (3 nm)/Co (1 nm)/Ta (3 nm) were grown on a thermally oxidized Si substrate by magnetron sputtering system with a pressure of 0.8 mTorr at room temperature. For the electrical transport measurement, the films with perpendicular magnetic anisotropy (PMA) were patterned into Hall bar with 10 $\mu$m width channel using photolithography and Ar ion milling techniques, as shown in Fig. 2(a). The anomalous Hall effect (AHE) measurements and logic operations were carried out using Keithley 2602B as the current source and Keithley 2182A as the nano voltmeter (Magnetic Field Probe Station, East Changing Technologies, China). All the measurements were carried out at room temperature.

Fig. 2(a) schematically illustrates two lateral interfaces at the center of Hall devices with local ion implantation. The pink area is implanted by the nitrogen (N) ion, where the implantation dose and energy of N ions are $3\times10^{13}$ cm$^{-2}$ and 10 KeV, respectively. The 0-degree (0D) and 45-degree (45D) lateral interfaces between implantation and non-implantation zones are indicated in Fig. 2(a). The 45D lateral interface introduces $\nabla H_K > 0$ along $y$ direction. The anomalous Hall resistance ($R_{Hall}$) is measured by sweeping the out-of-plane magnetic field ($H_Z$) under direct current $I_{DC}$ = +0.1 mA, as shown in Fig. 2(b), indicating that the Co layer has the PMA. And the two-step magnetic hysteresis loop confirms that the zones with ion implantation (without ion implantation) have small coercivity (large coercivity).

To investigate the influence of symmetry breaking and current-driven Néel wall motion mechanisms on the $M$, we measure hysteresis loops under $I_{DC}$ = 0, +1 and −1 mA, as shown in Fig. 3. One step, no-step and multiple steps hysteresis loops under $I_{DC}$ = 0, +1 and −1 mA is observed, respectively. To explore the reason for this phenomenon, we performed microscopic imaging of the magneto-optical Kerr effect (MOKE) on patterned Hall devices under different $I_{DC}$. Obviously, one step hysteresis loop under $I_{DC}$ = 0 mA (step ① in Fig. 3(a)) is induced by the DW pinning at the 45D interface. It is shown in the MOKE images that the DW keeps still at the 45D interface when $H_Z$ is switched from 3 Oe to 13.7 Oe (image <1> of Fig. 3(b)). The pinning effect at the 45D interface is resulted from the larger coercivity of the non-implanted zones. However, Fig. 3(d) shows the DW starts to cross the 45D interface coherently when the $H_Z$ is very small ($H_Z$ = 1.5 Oe), as shown in the image <2> of Fig. 3(d), suggesting no observable pinning at the 45D interface under $I_{DC}$ = +1 mA. Therefore, the positive current helps the DW to overcome the pinning effect at the 45D interface, leading to the coherent switching in Fig. 3(c). The $R_{Hall}$−$H$ loop under $I_{DC}$ = −1 mA (Fig. 3(e)) shows multiple steps (including the step ② at 10.5 Oe <$H_Z$ < 14 Oe, corresponding to the image <3> of Fig. 3(f)). The step ② is caused by the strong DW pinning at the 45D interface. Fig. 3(f) shows that the DW cannot cross the 45D interface when −18 Oe < $H_Z$ < 18 Oe. The switching at the right side of the 45D interface is not due to the DW across the 45D interface (The detailed MOKE images are presented in Fig. S1 of supplementary material). Thus, we can conclude that positive current favors DW to cross 45D interface, and negative current suppresses this process.

The different pinning effects at the 45D interface under different $I_{DC}$ can be explained by the current-driven Néel wall motion mechanism, as shown in Fig. 3(g). The $I_{DC}$ along +$x$ direction can produce an $H_{DL}$ on the left-handed Néel wall (proved in the *ref* 16) at the 45D interface. The $H_{DL}$ can drive the DW moving along the +$x$ direction so that it is easier for DW to overcome the pinning effect at the 45D interface. But the negative $I_{DC}$ produces the $H_{DL}$ that can drive the DW motion along the $-x$ direction. So, it is more difficult for the DW to cross the 45D interface. For the 0D interface, there is no obvious difference in the pinning field under various $I_{DC}$. As mentioned before, the $H_{DL}$ is zero for the 0D interface, and only $H_{eff(s)}$ affects the magnetic switching. This indicates that the DW motion is merely affected by the $H_{eff(s)}$ from symmetry breaking mechanism at the current amplitude of 1 mA. Therefore, the dominant mechanism is the current-driven Néel wall motion at the small current.

To study the contribution of the symmetry breaking and current-driven Néel wall motion mechanisms to DW motion at larger current, MOKE microscope images of Hall devices by injecting large different current pulses (pulse width of 50 ms) are measured, as shown in Fig. 4(a). When the current pulse is 18 mA, $+M$ ($-M$) domains are observed in the 45D (0D) interface, as shown in the image 1 of Fig. 4(a). The opposite $M$ switched by the same current at 45D and 0D interfaces is due to the opposite symmetry (reverse $\nabla H_K$) of the 45D and 0D interfaces. The $M$ is switched from $+M$ to $-M$ at the 45D interface and from $-M$ to $+M$ at the 0D interface when the current pulse is reversed to $-18$ mA (image 2), which is also due to the symmetry breaking mechanism. Therefore, we can observe the deterministic magnetic switching without an external magnetic field by the symmetry breaking mechanism at the large current.

To further explore the contribution of symmetry breaking and current-driven Néel wall motion mechanisms to DW motion within the current range between 1 mA to 18 mA, the DW motion under different currents is measured. We first initialized the $M$ by applying the initialization current of 18 mA in the $+x$ direction. Then the MOKE image is measured after each current pulse. The images of DW motion are selected and presented in Fig. 4(a) when decreasing the current pulse. The chirality of the DW in Fig. 4(a) at the 45D interface is illustrated in Fig. 4(b), where $H_{eff(S)}$ and $H_{DL}$ are in the competitive relationship with $\nabla H_K > 0$ along $y$ direction. The DW has not been changed when the current pulse is reduced from 18 to 8 mA, indicating that the DW cannot be driven by current pulses at large current. This means the symmetry breaking mechanism dominates at this current range. However, the DW starts to move in the $+x$ direction when lowering current pulse from 8 to 3 mA (from images 1-1 to 1-4). This is the DW motion direction by SOT due to the current-driven Néel wall motion mechanism. The $M$ is not determined by the symmetry breaking mechanism at the 45D interface under smaller current pulses. Further reducing the current pulse from 3 to 1 mA, the domain structure remains the same as the image 1-4. Therefore, when the current pulse decreases from 8 to 3 mA, the contribution of the symmetry breaking mechanism tends to decrease, but the contribution of the current-driven Néel wall motion mechanism tends to increase.

In order to verify the above conclusion, MOKE images are measured with the current pulse increasing from 1 mA to 18 mA after applying the initialization current of −18 mA. As shown in the domain structure of image 2 (initialized by a −18 mA current pulse), the −$M$ domains are distributed at the 45D interface. The current-driven Néel wall motion mechanism cannot change the $M$ at 45D interface due to the absence of DW. When the current pulse increases from 1 to 4 mA, this domain structure is unchanged (image 2-1), implying that the contribution of the symmetry breaking mechanism is very small. But when the current pulse is 4 mA, nucleation of a +$M$ domain occurs at the 45D interface (image 2-2) due to the symmetry breaking mechanism. And +$M$ domain gradually grows as the current pulse increases from 4 to 8 mA (images 2-3 and 2-4) due to the increasing contribution of the symmetry breaking mechanism and decreasing current-driven Néel wall motion mechanism. The domain structure of image 2-4 is maintained when the current pulse is in the range of 8 mA and 18 mA. Thus, when the current pulse rises from 3 to 8 mA, the contribution of the symmetry breaking mechanism gradually increases, and the current-driven Néel wall motion mechanism gradually decreases, verifying the previous conclusion. Thus, the contribution of symmetry breaking and current-driven Néel wall motion can be controlled by the current amplitude. The competition of the two mechanisms manipulated by current amplitude is confirmed at negative current, as shown in Fig. S2 of supplementary materials.

The reason for the two mechanisms contributing differently under small and large current is as follows: at a small current, the SOT that is exerted in the $M$ (pointed to $y$ axis) of the DW brings a perpendicular $H_{DL}$ ($\boldsymbol{y} \times \boldsymbol{x}$), resulting in the DW motion. However, under a large current pulse, the $M$ tends to the $y$ direction (the orientation of spin), which leads to the lower value of $H_{DL}$. Further increase the current, the $M$ all point to the $y$ axis, so that there is no domain wall in the device. In this case, the only mechanism to switch the $M$ is $H_{eff(s)}$ caused by symmetry breaking.

The competition of $H_{DL}$ and $H_{eff(s)}$ controlled by current amplitude at 45D lateral interface can lead to field-free non-hysteresis magnetization switching, as shown in Fig. 4(c) (Fig. 4(d)). Utilizing the non-hysteresis effect magnetic switching without an external magnetic field, we can realize non-resetting AND and OR Boolean logic gates without resetting (The concept of non-resetting function is described in supplementary S3). The input and output setups are illustrated in Fig. 2(a). Two currents $I_A$ and $I_B$, as inputs of logic, are applied along the $x$ axis simultaneously. Input values of "1" and "0" are represented by $I_{A/B}$ values 1.5 mA and 4 mA, respectively. The measured $R_{Hall}$ serves as the output, which can be identified as $R_{Hall} > -0.3$ Ω (logic "1") and $R_{Hall} < -0.28$ Ω (logic "0"). The truth table and experiment results of AND and OR logic gates are listed in Fig. 5(a) and Fig. 5(b). Thanks to the non-hysteresis effect behavior of magnetic switching, the AND and OR logic gates in this work successfully avoid resetting before each operation (non-resetting behavior). The non-resetting logic gates provide a possible way toward the application of energy-efficient and area-efficient spin logic.

In summary, the current-driven DW motion is observed at lateral interfaces by local N ion implantations. Due to the 45D lateral interface, both symmetry breaking and current-driven Néel wall motion mechanisms can affect the current-driven DW motion. The symmetry breaking (current-driven Néel wall motion) mechanism plays a major role at the large (small) current. In addition, the non-hysteresis effect magnetic switching is obtained by the competition of these two mechanisms. Utilizing the non-hysteresis effect magnetic switching, we can realize the non-resetting AND and OR logic gates. The realization of non-resetting logic gates will shed light on the application of energy-efficient and area-efficient spin logic.

**SUPPLEMENTARY MATERIAL**

See supplementary material for the detailed MOKE images, competition of the two mechanisms manipulated at negative current, and concept and advantages of non-resetting function.

**ACKNOWLEDGMENTS**

This work is supported by the Chinese Academy of Sciences, Grant No. XDA18000000, Y201926, and 2020118.

**CONFLICT OF INTEREST**

The authors have no conflicts to disclose.

**DATA AVAILABILITY**

The data that support the findings of this study are available from the corresponding author upon reasonable request.

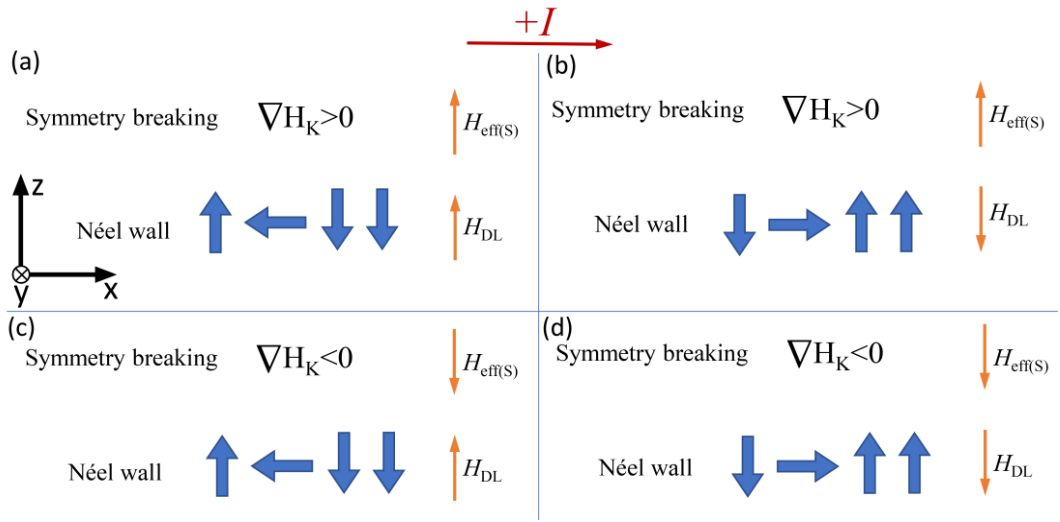

**Fig. 1.** The schematics of the effective fields caused by symmetry breaking and left-handed Néel-type DW motion when *M* of the DW pointes to (a) –*x* and (b) +*x* with $\nabla H_K > 0$, and *M* of the DW points to (c) –*x* and (d) +*x* with $\nabla H_K < 0$. The $\nabla H_K$ is the gradient anisotropy field along *y* axis. The $H_{eff(S)}$ is the effective field caused by symmetry breaking, and the $H_{DL}$ is the damping-like effective field from current-driven DW motion.

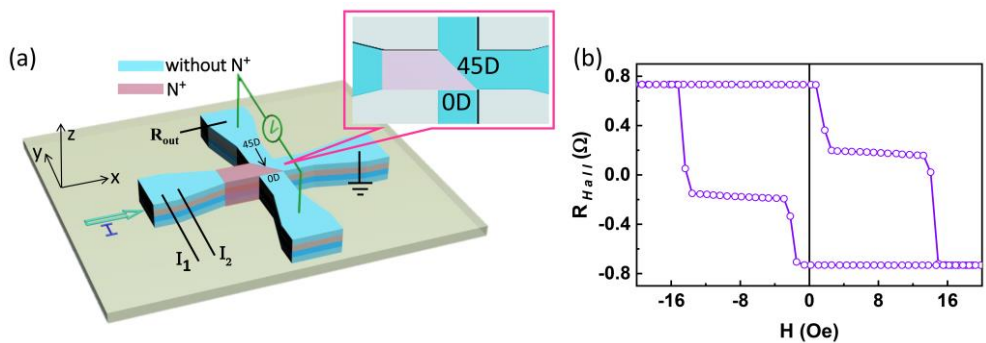

**Fig. 2.** (a) Hall device schematics, coordinate systems and the electrical measurement set-up. (b) The out-of-plane magnetic hysteresis of the device (a) measured under perpendicular magnetic fields.

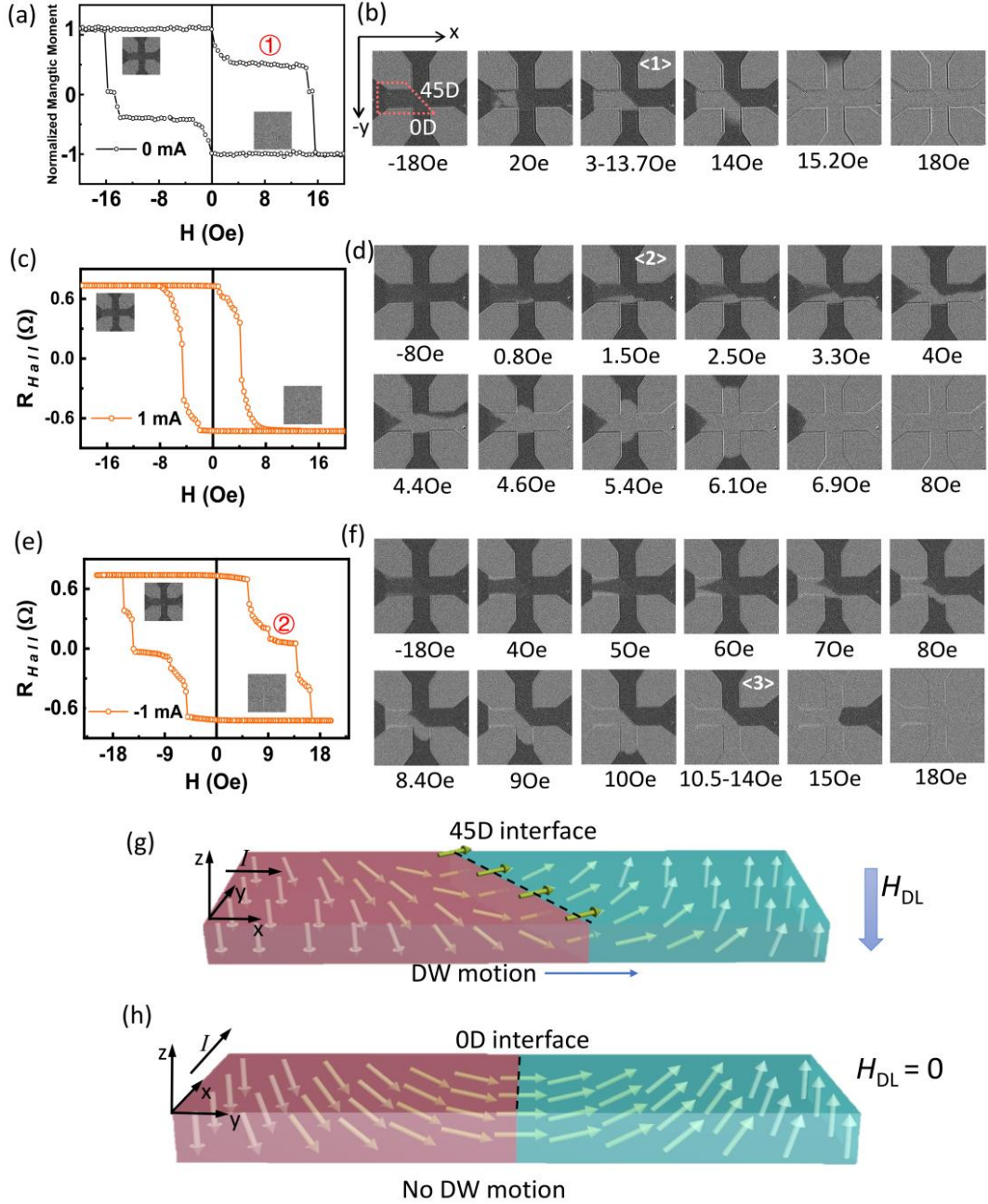

**Fig. 3.** (a-f) The out-of-plane magnetic hysteresis of the Hall device and the corresponding MOKE images measured under $I_{DC}$ = 0 mA (a,b), $I_{DC}$ = 1 mA (c,d), and $I_{DC}$ = −1mA (e,f), respectively. The black area in the images represents upward magnetization (+$M$), and the gray area represents downward magnetization (−$M$) within Hall devices areas. (g) and (h) The DW motion schematics and when the moment of the ion implantation region is down by current for the 45D (g) and 0D (h) interfaces.

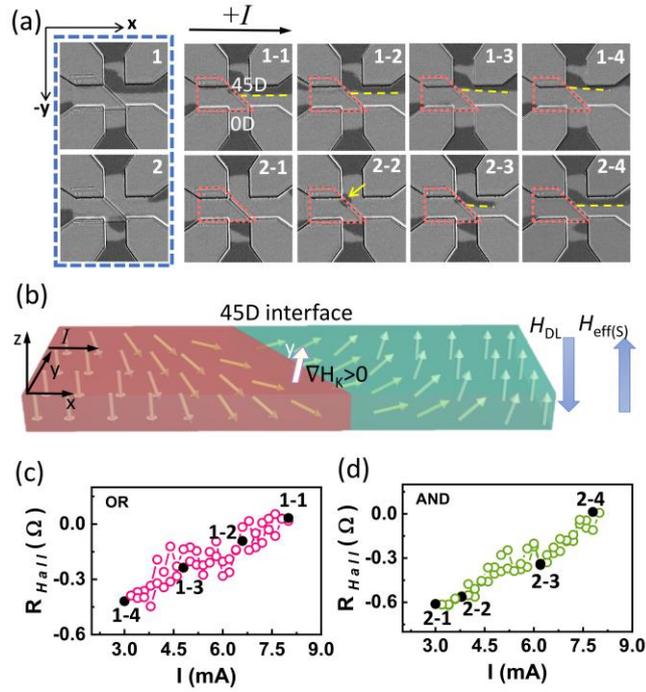

**Fig. 4.** (a) The MOKE images of the device with local ion implantation after a large initial current of 18 mA (image 1) and -18 mA (image 2). Images 1-1 to 1-4 are ones token after current pulses of 8, 6.2, 4.6 and 3 mA, and images 2-1 to 2-4 are captured after current pulses of 3, 4, 6.2 and 8 mA. The dashed yellow line and yellow arrows are used to mark the interface of different domains and the position of changing DW, respectively. (b) The magnetic domains schematic under the competition of $H_{DL}$ and $H_{eff(s)}$ at $\nabla H_K > 0$ along $y$ direction. (c) and (d) non-hysteresis effect magnetic switching after the current pulse of 18 mA (c) and −18 mA (d) when current changes from 3 to 8 mA.

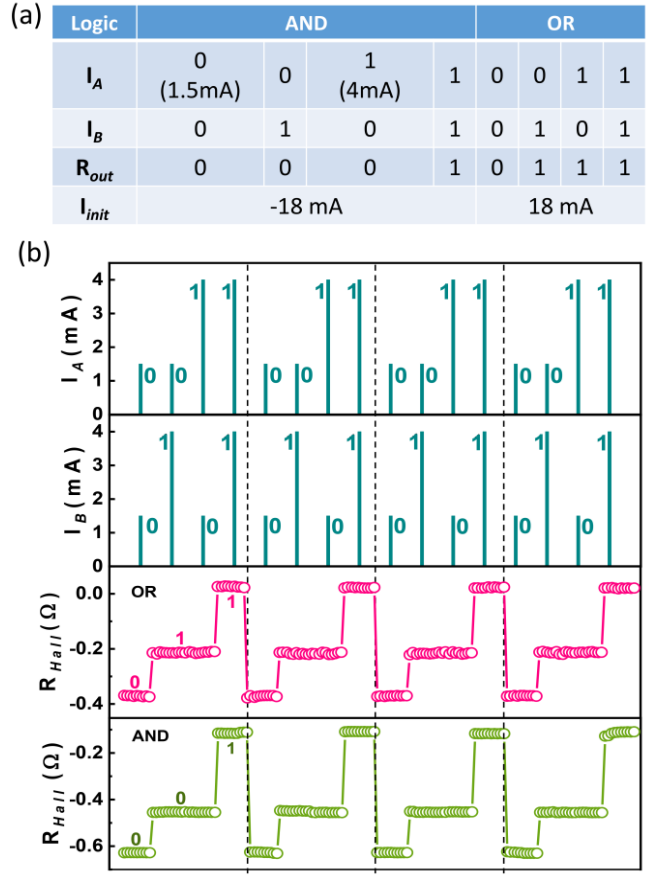

**Fig. 5.** (a) Truth table for OR and AND Boolean logic gates. (b) Demonstration of AND and OR logic gates.

# References

1. M. Yang, K. Cai, H. Ju, K. W. Edmonds, G. Yang, S. Liu, B. Li, B. Zhang, Y. Sheng, S. Wang, Y. Ji, and K. Wang, Sci. Rep. **6**, 20778 (2016).
2. I. M. Miron, K. Garello, G. Gaudin, P. J. Zermatten, M. V. Costache, S. Auffret, S. Bandiera, B. Rodmacq, A. Schuhl, and P. Gambardella, Nature **476**, 189 (2011).
3. L. Q. Liu, C. F. Pai, Y. Li, H. W. Tseng, D. C. Ralph, and R. A. Buhrman, Science **336**, 555 (2012).
4. L. Q. Liu, O. J. Lee, T. J. Gudmundsen, D. C. Ralph, and R. A. Buhrman, Phys. Rev. Lett. **109**, 5 (2012).
5. Y. Li, J. Liang, H. Yang, H. Zheng, and K. Wang, Appl. Phys. Lett. **117**, 092404 (2020).
6. Y. Cao, A. Rushforth, Y. Sheng, H. Zheng, and K. Wang, Adv. Funct. Mater. **29**, 1808104 (2019).
7. B. Cui, D. Li, J. Yun, Y. Zuo, X. Guo, K. Wu, X. Zhang, Y. Wang, L. Xi, and D. Xue, Phys.

Chem. Chem. Phys. **20**, 9904 (2018).

8. J. Zhou, T. Zhao, X. Shu, L. Liu, W. Lin, S. Chen, S. Shi, X. Yan, X. Liu, and J. Chen, Adv. Mater. **33**, e2103672 (2021).
9. I. M. Miron, T. Moore, H. Szambolics, L. D. Budaprejbeanu, S. Auffret, B. Rodmacq, S. Pizzini, J. Vogel, M. Bonfim, and A. Schuhl, Nat. Mater. **10**, 419 (2011).
10. N. Sato, F. Xue, R. M. White, C. Bi, and S. X. Wang, Nature Electronics **1**, 508 (2018).
11. Z. Luo, A. Hrabec, T. P. Dao, G. Sala, S. Finizio, J. Feng, S. Mayr, J. Raabe, P. Gambardella, and L. J. Heyderman, Nature **579**, 214 (2020).
12. Z. Luo, S. Schären, A. Hrabec, T. P. Dao, G. Sala, S. Finizio, J. Feng, S. Mayr, J. Raabe, P. Gambardella, and L. J. Heyderman, Phys. Rev. Appl. **15**, 034077 (2021).
13. Z. Luo, T. P. Dao, A. Hrabec, J. Vijayakumar, A. Kleibert, M. Baumgartner, E. Kirk, J. Cui, T. Savchenko, G. Krishnaswamy, L. J. Heyderman, and P. Gambardella, Science **363**, 1435 (2019).
14. Y. Cao, Y. Sheng, K. W. Edmonds, Y. Ji, H. Zheng, and K. Wang, Adv. Mater. **32**, e1907929 (2020).
15. N. Zhang, Y. Cao, Y. Li, A. W. Rushforth, Y. Ji, H. Zheng, and K. Wang, Adv. Electron. Mater. **6**, 2000296 (2020).
16. M. Yang, Y. Li, J. Luo, Y. Deng, N. Zhang, X. Zhang, S. Li, Y. Cui, P. Yu, T. Yang, Y. Sheng, S. Wang, J. Xu, C. Zhao, and K. Wang, Phys. Rev. Appl. **15**, 054013 (2021).
17. G. Q. Yu, P. Upadhyaya, Y. B. Fan, J. G. Alzate, W. J. Jiang, K. L. Wong, S. Takei, S. A. Bender, L. T. Chang, Y. Jiang, M. R. Lang, J. S. Tang, Y. Wang, Y. Tserkovnyak, P. K. Amiri, and K. L. Wang, Nat. Nanotechnol. **9**, 548 (2014).
18. S. Chen, J. Yu, Q. Xie, X. Zhang, W. Lin, L. Liu, J. Zhou, X. Shu, R. Guo, Z. Zhang, and J. Chen, ACS Appl. Mater. Interfaces **11**, 30446 (2019).
19. L. You, O. Lee, D. Bhowmik, D. Labanowski, J. Hong, J. Bokor, and S. Salahuddin, Proc. Natl. Acad. Sci. U. S. A. **112**, 10310 (2015).
20. H. Wu, J. Nance, S. A. Razavi, D. Lujan, B. Dai, Y. Liu, H. He, B. Cui, D. Wu, K. Wong, K. Sobotkiewich, X. Li, G. P. Carman, and K. L. Wang, Nano. Lett. **21**, 515 (2021).
21. A. V. Khvalkovskiy, V. Cros, D. Apalkov, V. Nikitin, M. Krounbi, K. A. Zvezdin, A. Anane, J. Grollier, and A. Fert, Phys. Rev. B **87**, 020402 (2013).
22. K. S. Ryu, L. Thomas, S. H. Yang, and S. Parkin, Nat Nanotechnol **8**, 527 (2013).

**Current controlled non-hysteresis magnetic switching in the absence of magnetic field**

Yanru Li[1,2], Meiyin Yang[1,2]*, Guoqiang Yu[3,4], Baoshan Cui[3,4], and Jun Luo[1,2]*

[1]*Key Laboratory of Microelectronic Devices and Integrated Technology, Institute of Microelectronics, Chinese Academy of Sciences (IMECAS), Beijing 100029, China*
[2]*University of Chinese Academy of Sciences (UCAS), Beijing 100049, China*
[3]*Songshan Lake Materials Laboratory, Dongguan, Guangdong 523808, China*
[4]*Institute of Physics, Chinese Academy of Sciences, Beijing 100190, China*

E-mail: luojun@ime.ac.cn, yangmeiyin@ime.ac.cn

**Supplementary Information**

Table of Contents:
S1. The DW motion under an out-of-plane magnetic field at $I_{DC}$ = -1 mA.
S2. The contribution of symmetry breaking and current-driven Néel wall motion mechanisms to DW motion when the current pulse range is from -11 mA to -18 mA.
S3. The concept and advantages of non-resetting function.

**S1. The DW motion under an out-of-plane magnetic field at $I_{DC}$ = -1 mA.**

When the out-of-plane magnetic field increases from 13.5 to 18 Oe, the DW motion at $I_{DC}$ = -1 mA is shown in Fig. S1. The DW at the 45D interface cannot cross the 45D interface even if the magnetic field increases. In other words, the DW at the 45D interface is strongly pinned. But the DW at pad 2 can move to $x$ direction when 13.5 Oe < $H_Z$ < 15 Oe, and the DW at pad 3 moves to -$y$ direction when 15 Oe < $H_Z$ < 18 Oe, leading to the complete -$M$ domains. The DW motion is similar to the above phenomenon when the $H_Z$ changes from -13.5 Oe to -18 Oe. So, the switching can also reach saturation even if the DW cannot cross the 45D interface.

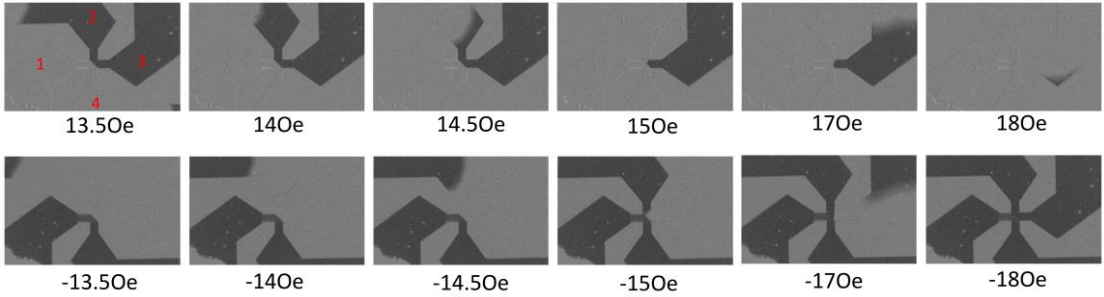

**Fig. S1.** The MOKE images of the magnetic domain for -1 mA. The black area in the images represents upward magnetization (+$M$), and the gray area represents downward magnetization (-$M$) within Hall devices areas.

**S2. The contribution of symmetry breaking and current-driven Néel wall motion mechanisms to DW motion when current pulse range is from -11 mA to -18 mA**

We also measure MOKE images when the current pulse range is from -11 mA to -18 mA. After being initialized by 18 mA (image 1), the MOKE images pictured after selected current pulses are present in Fig. S2 under increasing negative current pulse. The current-driven Néel wall motion mechanism cannot change the domain structure of image 1. The reason is that the DW is pinned at the 45D interface even if the DW moves in the -$y$ direction under negative current. The domain structure of image 1 is unchanged when the negative current pulse increases from -11 mA to -12 mA, meaning that the contribution of the symmetry breaking mechanism is very small. The nucleation occurs near the 45D interface when the current pulse is -12 mA (image S1-1) due to the symmetry breaking mechanism. As the negative current pulse increase from the -12 mA to -17 mA, the -$M$ domains gradually grow until +$M$ domains disappear at the 45D interface (from image S1-2 to image S1-4), indicating the increasing contribution of the symmetry breaking mechanism and decreasing current-driven Néel wall motion mechanism. The domain structure of image S1-4 is unchanged when the current pulse range is -18 mA and -17 mA. So, when the current pulse rises from -11 to -18 mA, the contribution of the symmetry breaking mechanism gradually increases, and the current-driven Néel wall motion mechanism gradually decreases.

The MOKE image after the initialization current of -18 mA is shown in image 2 of Fig. S2. This domain structure is maintained when the current pulse changes from -18 mA to -16 mA due to the dominant mechanism of symmetry breaking (image S2-1). But the DW moves in the -$y$ direction when the current pulse is -16 mA to -12 mA (from image S2-2 to image S2-4) due to the decreasing contribution of the symmetry breaking mechanism and increasing current-driven Néel wall motion mechanism. The domain structure of image S2-4 is unchanged when the negative current decreases from -12 mA to -11 mA. Thus, when the negative $I_{CP}$ decreases from -18 to -11 mA, the contribution of the symmetry breaking mechanism tends to decrease, but the contribution of the current-driven Néel wall motion mechanism tends to increase.

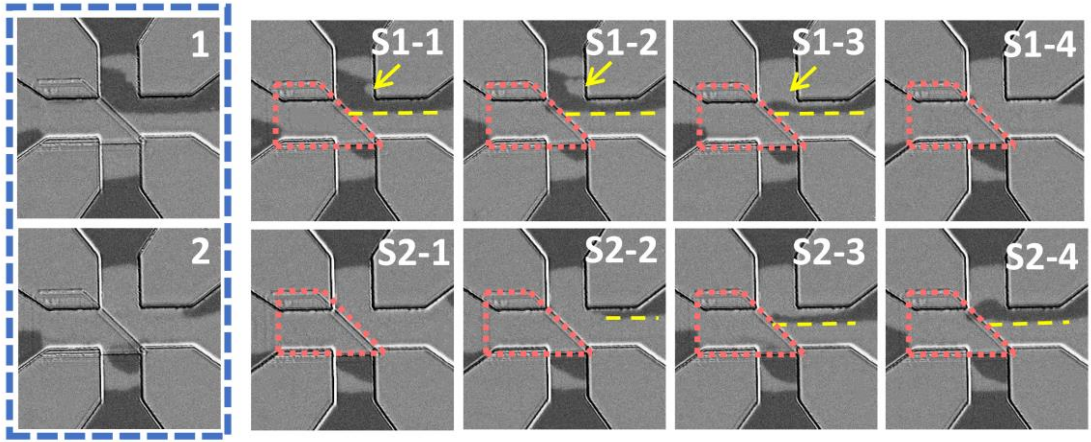

**Fig. S2.** The MOKE images of the device with local ion implantation after a large initial current, the dashed yellow line, and the yellow arrow are used to mark the interface of different domains and the position of changing domains, respectively.

**S3. The concept and advantages of non-resetting function.**

Fig. S3(a) shows the magnetic switching loop with the hysteresis effect, which means that one input corresponds to two output levels. The output is highly related to the previous state of the devices. To guarantee that one input results in only one output, the device's initial state is set to be the same. This is the reason why magnetic switching with the hysteresis effect behavior requires resetting before every operation. For example, the logic gate of AND can be realized when the output is from state 1 to 2. After the AND operation, a sufficiently large initializing current is required to return to state 1 before the next AND operation. Without resetting, the output will always be at state 2 regardless of the input.

Fig. S3(b) represents the magnetic switching loop without hysteresis effect. Here, in the working range, one input value only results in one output state. Thus, the resetting procedure is avoided.

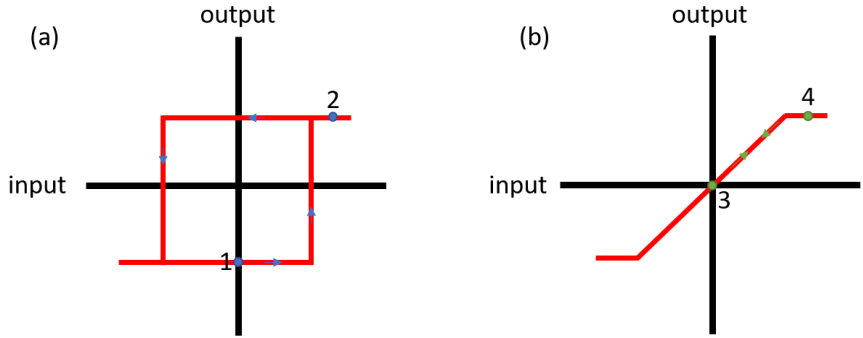

**Fig. S3**. (a) magnetic switching with hysteresis effect. (b) non-hysteresis magnetic switching.

To compare the logic gates realized by magnetic switching with and without hysteresis effect, we present the working mode for the AND logic gate, as is shown in Fig. S4. Because the AND logic by non-linear switching is reset before every operation, more time and energy consumption are consumed. The non-resetting logic gate using the non-hysteresis magnetic switching can save almost 50% time and power compared with the resetting logic.

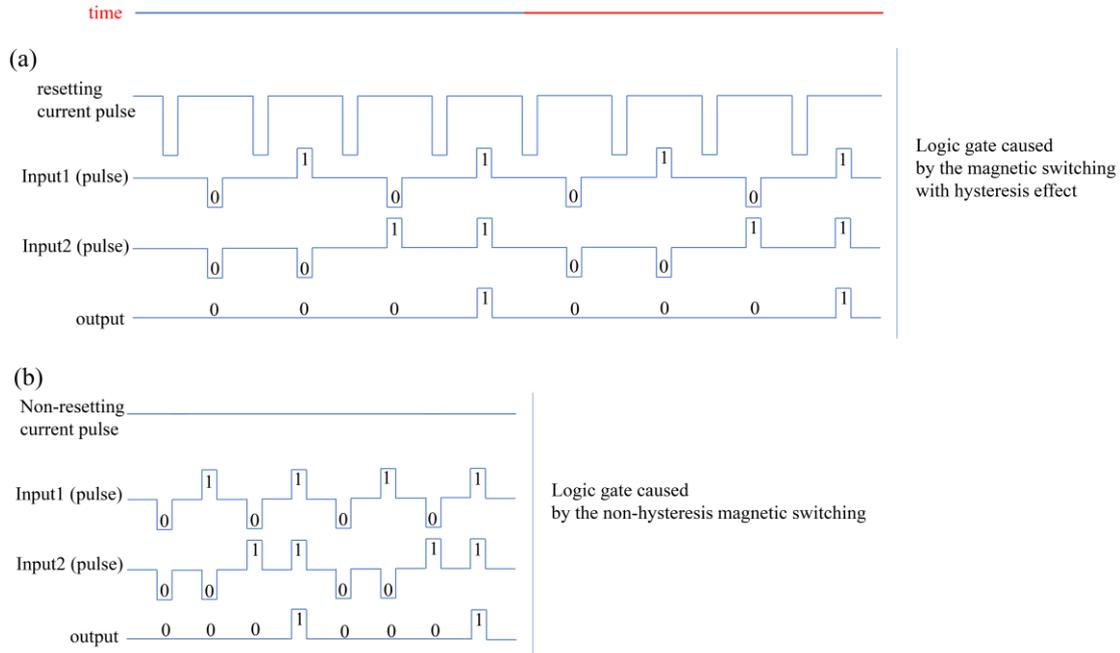

**Fig. S4**. The AND logic gate caused by (a) magnetic switching with hysteresis effect (b) non-hysteresis magnetic switching